\documentclass[showpacs,preprintnumbers,amsmath,amssymb, nofootinbib]{revtex4}
\usepackage{fontenc}
\usepackage{graphicx}

\usepackage[dvips]{hyperref}

\usepackage{amsmath}

\begin{document}

\title{Comments on absorption cross section for Chern-Simons black holes in five dimensions}
\author{P.A. Gonz\'{a}lez}
\email{pgonzalezm@ucentral.cl} \affiliation{ Escuela de Ingenier\'{i}a Civil en Obras Civiles. Facultad de Ciencias F\'{i}sicas y Matem\'{a}ticas, Universidad Central de Chile, Avenida Santa Isabel 1186, Santiago, Chile}
\affiliation{Universidad Diego Portales, Casilla 298-V, Santiago,
Chile.}
\author{Joel Saavedra}
\email{joel.saavedra@ucv.cl} \affiliation{Instituto de
F\'{\i}sica, Pontificia Universidad Cat\'olica de Valpara\'{\i}so,
Casilla 4950, Valpara\'{\i}so, Chile.}
\date{\today}
\vspace{5cm}

\begin{abstract}

In this paper we study the effects of  black hole mass on the
absorption cross section for a massive scalar field propagating in
a $5$-dimensional topological Chern-Simons black hole
at the low-frequency limit. We consider the two branches of black
hole solutions $(\alpha=\pm 1)$ and we show that, if the mass of
black hole increase the absorption cross section decreases at the
zero-frequency limit for the branch $\alpha=-1$ and for the other
branch, $\alpha=1$, the behavior is opposite, if the black hole
mass increase the absorption cross section increases. Also we find that beyond a certain frequency value, the
mass black hole does not affect  the absorption cross section.
\end{abstract}

\maketitle

\section{Introduction}
In general relativity the field equation adopt the following form
\begin{equation}
G_{\mu \nu }=8\pi G_{N}T_{\mu \nu }~,  \label{einsteineq}
\end{equation}
where $G$ is a symmetric tensor that depends of the metric
$g_{\mu\,\nu}$ and their first derivatives, furthers its
divergences is vanished
\begin{equation}
G_{;\mu }^{\mu \nu }=0~.  \label{divergence}
\end{equation}
In four dimensions these conditions allow to determinate the
Einstein tensor, at least of arbitrary multiplicative constant
factor
\begin{equation}
G_{\mu \nu} =R_{\mu \nu }-\frac{1}{2}g_{\mu \nu }R-\Lambda g_{\mu \nu
}~. \label{einsteintensor}
\end{equation}
The Einstein tensor is the only symmetric and conserved tensor
depending on the metric and its derivatives, which is linear in
the second derivatives of the metric. The invariant action that
arise these fields equations, it is the Einstein-Hilbert action
with cosmological constant ($\Lambda$). In higher dimensions, the
potential problem is to find the most general action that arise
a set of second order field equations. The solution to this
problems it is the called Lanczos-Lovelock (LL) action \cite{LL}. This
action is non linear in the Riemann tensor, and differs from the
Einstein-Hilbert action only if the space-time has more than $4$
dimensions. Therefore, the Lanczos-Lovelock action is the most
natural extension of general relativity in higher dimensional
space-times, that  generate second order field equations.
This action in $d$ dimensions can be written as follow
\begin{equation}
S=\sum_{2p<d}\alpha _{p}S_{p}~,  \label{action1}
\end{equation}
with
\begin{equation}
S_{p}=\frac{1}{2p!}\int \sqrt{-g}\,\delta _{[\alpha _{1}\cdots
\alpha _{2p}]}^{[\beta _{1}\cdots \beta _{2p}]}\,R_{\beta
_{1}\beta _{2}}^{\alpha _{1}\alpha _{2}}\cdots
R_{\beta_{2p-1}\beta _{2p}}^{\alpha _{2p-1}\alpha _{2p}}\,d^{d}x~,
\label{actiond}
\end{equation}
 where the $\alpha_p$ are arbitrary coefficients with dimension of
 $mass^{d-2p}$ and the symbols $\delta_{[\cdots]}^{[\cdots]}$ are
 define by
 \begin{equation}
\delta _{[\alpha _{1}\cdots \alpha _{2p}]}^{[\beta _{1}\cdots
\beta _{2p}]}=\left|
\begin{array}{lll}
\delta _{\alpha _{1}}^{\beta _{1}} & \cdots  & \delta _{\alpha
_{1}}^{\beta
_{2p}} \\
\vdots  & \ddots  & \vdots  \\
\delta _{\alpha _{2p}}^{\beta _{1}} & \cdots  & \delta _{\alpha
_{2p}}^{\beta _{2p}}
\end{array}
\right|~,\label{delta}
\end{equation}
it represents the generalizes delta Kronecker.

In the particular case when $d=4$, we can identify constants
$\alpha_0=\frac{\lambda}{8\,\pi\,G_N}$ and
$\alpha_1=\frac{1}{8\,\pi\,G_N}$ i.e are related to the
cosmological constant and gravitational Newton constant
respectively.

On the other hand, differential forms allows to characterize  LL
action and those dynamical equations. In order to do this task we
introduce  the vielbein $e^a$ and the spin connection
$\omega^{ab}$  1-forms. These 1-forms are related to curvature and torsion
two-forms as follow $R^{ab}=d\omega^{ab}+\omega^a_c\omega^{cb}$
and $T^a=de^a+\omega^a_be^b$. So, the lagragian can be written as 
\begin{equation}\label{accionLL}\
L_{LL}= \int \sum_{q=0}^{[d/2]}\alpha_qL^q~,
\end{equation}%
where $\alpha_q$ are arbitrary constants and
\begin{equation}\label{polinomio}\
L^{q}=\epsilon_{a_{1}...a_{d}}R^{a_{1}a_{2}}...R^{a_{2q-1}a_{2q}}e^{a_{2q+1}}...e^{a_{d}}.
\end{equation}%

In generic form the coefficients are arbitrary in the LL action.
However, they can be fixed in order to have a unique vacuum or a
unique cosmological constant. In this case the coefficient are
given by $\alpha_q:=c_{q}^{k}$ where
$c_{q}^{k}=\frac{l^{2(q-k)}}{d-2q}(^{k}_{q})$ for $q\leq k$ and
vanished for $q>k$, which $1\leq k\leq [\frac{d-1}{2}]$. Then, the
action can be written as follows
\begin{equation}\label{accionk}\
I_{k}= \kappa\int \sum_{q=0}^{[k]}c_q^kL^q~.
\end{equation}%
This action possesses two fundamentals constants $\kappa$ (related
to Newton gravitational constant $G_{k}$) and $l$, (related to
cosmological constant $\Lambda$).
\begin{equation}\label{definitionk}\
\kappa=\frac{1}{2(d-2)!\Omega _{d-2}G_{k}}~,
\end{equation}%
\begin{equation}\label{lambda}\
\Lambda=-\frac{(d-1)(d-2)}{2l^2}~,
\end{equation}%
where $\Omega_{d-2}$ is the volume of a $(d-2)$-unit sphere. For $k=1$  the Einstein- Hilbert action is recovered. It is
worth nothing that in odd dimensions the theory is gauge invariant
under the (anti-)de Sitter or the Poincare groups and the
lagrangian is a Chern-Simons form.  

Chern-Simons black holes are special solutions of
gravity theories in higher odd dimensions which contain
higher powers of curvature. These theories are consistent
Lanczos-Lovelock theories resulting in second order field
equations for the metric with well defined AdS asymptotic
solutions. For spherically symmetric topologies, these black holes
are labelled by an integer $k$ which specifies the higher order of
curvature present in the Lanczos-Lovelock action and it is related
to the dimensionality $d$ of spacetime by the relation
$d-2k=1$~\cite{Crisostomo:2000bb}. If $d-2k=1$, the solutions are
known as Chern-Simons black holes (for a review on the
Chern-Simons theories see \cite{Zanelli:2005sa}). These solutions
were further generalized to other topologies~\cite{Aros:2000ij}
and they can be described in general by a non-trivial transverse
spatial section $\sum_{\gamma}$ of $(d-2)$-dimensions labelled by
the constant $\gamma=+1, -1, 0$ that represents the curvature of
the transverse section, corresponding to a spherical, hyperbolic
or plane section, respectively. The solution describing a black
hole in a free torsion theory can be written as~\cite{Aros:2000ij}
\begin{equation}\label{dtopologicalmetric}\
ds^{2}=-\Big{(} \gamma +\frac{r^{2}}{l^{2}}-\alpha \Big{(} \frac{2G_{k}\mu }{%
r^{d-2k-1}}\Big{)} ^{\frac{1}{k}}\Big{)} dt^{2}+\frac{dr^{2}}{\Big{(} \gamma +%
\frac{r^{2}}{l^{2}}-\alpha \Big{(} \frac{2G_{k}\mu }{r^{d-2k-1}}\Big{)} ^{%
\frac{1}{k}}\Big{)} }+r^{2}d\sigma _{\gamma }^{2}~,
\end{equation}%
where $\alpha =(\pm 1)^{k+1}$ and the constant $\mu$ is related to
the horizon $r_{+}$ through
\begin{equation}\label{relationr}\
\mu =\frac{r_{+}^{d-2k-1}}{2G_{k}}(\gamma
+\frac{r_{+}^{2}}{l^{2}})^{k}~,
\end{equation}%
and to the mass $M$ by
\begin{equation}\label{relationM}\
\mu =\frac{\Omega _{d-2}}{\Sigma _{d-2}}M+\frac{1}{2G_{k}}\delta
_{d-2k,\gamma }~,
\end{equation}%
here $\Sigma _{d-2}$ denotes the volume of the transverse space.
As can be seen in (\ref{dtopologicalmetric}), if $d-2k\neq1$ the
$k$ root makes the curvature singularity milder than the
corresponding black hole of the same mass. At the exact
Chern-Simons limit $d-2k=1$, the solution has similar structure
like the (2+1)-dimensional BTZ black hole with a string-like
singularity. We are merely interested here for the hyperbolic
topology with $\gamma=-1$. In this case  $d\sigma^{2}_{-1}$ in
(\ref{dtopologicalmetric}) is the line element of the
$(d-2)$-dimensional manifold $\sum_{-1}$, which is locally
isomorphic to the hyperbolic manifold $H^{d-2}$.

The spherical topologies  Chern-Simons black holes have similar
causal structure as the (2+1)-dimensional BTZ black
hole~\cite{Banados:1992wn}, and they have positive specific heat
and therefore are thermodynamical stable. For hyperbolic
topologies, the Chern-Simons black holes resemble to the
topological black holes~\cite{topological} in their zero mass
limit, and their thermodynamic behavior was studied
in~\cite{Aros:2000ij} and the quasi-normal modes (QNMs), the
reflection coefficients, the transmission coefficients and the
absorption cross section of scalar perturbations was studied
in~\cite{Gonzalez:2010vv}. The QNMs of the Chern-Simons black
holes depends on the black hole parameters and on the fundamental
constants of the system, also it depends on the curvature
parameter $k$. Besides this curvature parameter is related to the
number of symmetries in the theory, taking his maximum value for
Chern-Simons theories.  Also, at low frequency limit that there is
a range of modes with high angular momentum which contributes at
the absorption cross section. An illustrative  example of Chern-Simons black holes is provided by
the Gauss-Bonnet theory for $d=5$ and $k=2$. Static local
solutions of this theory are well studied over the
years~\cite{gbpapers}. This theory has two branches of solutions.
If there is a fine tuning between $k$ and the Gauss-Bonnet
coupling constant $\alpha$, the two solutions coincide to the
Chern-Simons black hole solution which has maximum symmetry. This
is known as the Chern-Simons limit (for a review
see~\cite{Charmousis:2008kc}). The stability of these solutions
has also been studied~\cite{stability}. It was found
in Ref.~\cite{Charmousis:2008ce} that one of these solutions suffers
from ghost-like instability up to the strongly coupled
Chern-Simons limit where linear perturbation theory breaks down.

The Hawking radiation is a semiclassical effect and it gives the
thermal radiation emitted by a black hole.  At the event horizon,
the Hawking radiation is in fact blackbody radiation. However,
this radiation still has to traverse a non-trivial curved
spacetime geometry before it reaches a distant
 observer who  detects it. The surrounding spacetime
thus works as a potential barrier for the radiation giving a
deviation from the blackbody radiation spectrum, seen
by an asymptotic observer~\cite{Maldacena:1996ix,Harmark:2007jy}. So, the total flux observed at infinity is that of a d-dimensional greybody at the
Hawking temperature.

In the present work, we focus our study on the effect of the mass of topological Chern-Simons black hole  in five dimensions on
the absorption cross section for scalar fields and mainly we show that from a certain frequency the mass
black hole mass does not affect at the absorption cross section.


\section{Quasinormal modes and absorption cross sections of scalar perturbations}

The metric of Chern-Simons black hole under consideration\footnote{See Ref.~\cite{Gonzalez:2010vv} for details and discussions.} is
\begin{equation}\label{metric2}\
ds^{2}=-f(r)dt^{2}+\frac{1}{f(r)}dr^{2}+r^{2}d\sigma
_{\gamma}^{2}~,
\end{equation}
where,
\begin{equation}
f(r)=\gamma +\frac{r^{2}}{l^{2}}-\alpha(2\mu
G_{k})^{\frac{1}{k}}~,
\end{equation}
and the horizon is located at
\begin{equation}\label{horizon}\
r_{+}=l\sqrt{\alpha(2\mu G_{k})^{\frac{1}{k}}-\gamma}~.
\end{equation}
Considering the horizon geometry with a negative curvature
constant, $\gamma=-1$, the allowed range of $\mu$ for $r_+\geq0$
are: if $k$ is odd, $\alpha=1$, $\mu\geq\frac{-1}{2G_k}$; if $k$
is even and $\alpha=1$, $\mu \geq0$ and if $k$ is even and
$\alpha=-1$, $\frac{1}{2G_k}\geq\mu\geq0$ \cite{Aros:2000ij}.

Performing the change of variables $v=1-l^2p/r^2$, where
$p=1+\alpha(2\mu G_{k})^{\frac{1}{k}}$ and $t=lt$, and written the
Klein-Gordon equation for a minimally coupled scalar field in the
background of a Chern-Simons black hole (in d-dimensions) we
obtain
\begin{equation}\label{radial}\
v(1-v)\partial_{v}^2R(v)+\left[1+\left(\frac{d-5}{2}\right)v\right]\partial_{v}R(v)+
\left[\frac{\omega^2}{4pv}-\frac{Q}{4p}-\frac{m^2l^2}{4(1-v)}\right]R(v)=0~,
\end{equation}
where $Q$ corresponds to eingenvalues for Laplace operator in the
hyperbolic manifold $\sum_{d-2}$, and it is given by
$Q=\left(\frac{d-3}{2}\right)^2+\xi^2$, without any
identifications of the pseudosphere the spectrum of the angular
wave equation is continuous, thus $\xi$ takes any real value
$\xi\geq0$. Since the $(d-2)-$dimensional manifold $\sum$ is a
quotient space of the form $H^{d-2}/\Gamma$ and it is a compact
space of constant negative curvature, the spectrum of the angular
wave equation is discretized and thus $\xi$ takes discrete real
values $\xi\geq0$~\cite{Balazs:1986uj}. Now, the radial function
$R(v)$ becomes under the decomposition $R(v)=v^\alpha(1-v)^\beta
K(v)$, Eq.~(\ref{radial}) can be written as a hypergeometric
equation for K
\begin{equation}\label{hypergeometric}\
v(1-v)K''(v)+\left[c-(1+a+b)v\right]K'(v)-ab K(v)=0~.
\end{equation}
Where the coefficients are given by
\begin{equation}\label{a}\
a=-\left(\frac{d-3}{4}\right)+\alpha+\beta+\frac{i}{2}\sqrt{\frac{\xi^2}{p}+\left(\frac{d-3}{2}\right)^2\left(\frac{1}{p}-1\right)}~,
\end{equation}
\begin{equation}
b=-\left(\frac{d-3}{4}\right)+\alpha+\beta-\frac{i}{2}\sqrt{\frac{\xi^2}{p}+\left(\frac{d-3}{2}\right)^2\left(\frac{1}{p}-1\right)}~,
\end{equation}
\begin{equation}
c=1+2\alpha~,
\end{equation}
where $c$ cannot be an integer and the exponents $\alpha$ and
$\beta$ are
\begin{equation}
\alpha=\pm\frac{i\omega\sqrt{p}}{2p}~,
\end{equation}
\begin{equation}
\beta=\beta_\pm=\left(\frac{d-1}{4}\right)\pm\frac{1}{2}\sqrt{\left(\frac{d-1}{2}\right)^2+m^2l^2}~.
\end{equation}
Without loss of generality, we choose the negative signs for
$\alpha$. The general solution of Eq.~(\ref{hypergeometric}) takes
the form
\begin{equation}
K=C_{1}F_{1}(a,b,c;v)+C_2v^{1-c}F_{1}(a-c+1,b-c+1,2-c;v)~,
\end{equation}
which has three regular singular point at $v=0$, $v=1$ and
$v=\infty$. Here, $F_{1}(a,b,c;v)$ is a hypergeometric function
and $C_{1}$, $C_{2}$ are constants. Then, the solution for the
radial function $R(v)$ is
\begin{equation}\label{RV}\
R(v)=C_{1}v^\alpha(1-v)^\beta
F_{1}(a,b,c;v)+C_2v^{-\alpha}(1-v)^\beta
F_{1}(a-c+1,b-c+1,2-c;v)~.
\end{equation}

To obtain an exact expression for the quasi-normal modes of scalar
perturbations of a Chern-Simons black hole in $d-$dimensions we
need to impose boundary conditions on asymptotically AdS
spacetime. First, we have to impose our boundary conditions on the horizon that there exist only ingoing waves. This fixes  $C_2=0$. Then the radial
solution becomes
\begin{equation}\label{horizonsolutiond}
R(v)=C_1 e^{\alpha \ln v}(1-v)^\beta F_{1}(a,b,c;v)=C_1
e^{-i\omega\frac{\sqrt{p}}{2p}\ln v}(1-v)^\beta F_{1}(a,b,c;v)~.
\end{equation}
In order to implement boundary conditions at infinity ($v=1$), we
shall apply in Eq.~(\ref{horizonsolutiond}) the Kummer's formula
for the hypergeometric function \cite{M. Abramowitz}, with this
expression the radial function results in
\begin{eqnarray}\label{R}\
\nonumber R(v) &=& C_1 e^{-i\omega\frac{\sqrt{p}}{2p}\ln v}(1-v)^\beta\frac{\Gamma(c)\Gamma(c-a-b)}{\Gamma(c-a)\Gamma(c-b)} F_1(a,b,a+b-c,1-v)\\
&& +C_1 e^{-i\omega\frac{\sqrt{p}}{2p}\ln
v}(1-v)^{c-a-b+\beta}\frac{\Gamma(c)\Gamma(a+b-c)}{\Gamma(a)\Gamma(b)}F_1(c-a,c-b,c-a-b+1,1-v)~.
\end{eqnarray}
The flux is given by
$\textit{F}=\frac{\sqrt{-g}g^{rr}}{2i}\left(\varphi^{\ast}\partial_{r}\varphi-\varphi\partial_{r}\varphi^{\ast}\right)$,
and demanding that it vanishes at infinity we are able to
determined the two set of quasinormal modes
\begin{equation}\label{qnm}\
\omega=\mp\sqrt{\xi^2+\left(\frac{d-3}{2}\right)^2\left(1-p\right)}-i\sqrt{p}\left(2n+1\pm\sqrt{\left(\frac{d-1}{2}\right)^2+m^2l^2}\right)~,
\end{equation}
for $\beta_+$ and $\beta_-$, respectively.

On the other hand, the reflection and the transmission
coefficients are defined by
\begin{equation}\label{reflectiond}\
\Re :=\left|\frac{F_{\mbox{\tiny asymp}}^{\mbox{\tiny
out}}}{F_{\mbox{\tiny asymp}}^{\mbox{\tiny in}}}\right|, \qquad
\mbox{and} \qquad  \mathfrak{U}:=\left|\frac{F_{\mbox{\tiny
hor}}^{\mbox{\tiny in}}}{F_{\mbox{\tiny asymp}}^{\mbox{\tiny
in}}}\right|,
\end{equation}
so we need to know the behavior of the radial function both at the
horizon and at the asymptotic infinity. The behavior at the
horizon is given by Eq.~(\ref{horizonsolutiond}). Thus, to obtain
the asymptotic of the $R(v)$ we use $1-v=\frac{l^2p}{r^2}$ and
taking into account the limit of $R(v)$, Eq.~(\ref{RV}), when
$v\rightarrow1$, we have
\begin{equation}\label{Rinfinity}\
R(r) =
C_1\left(\frac{l\sqrt{p}}{r}\right)^{2\beta}\frac{\Gamma(c)\Gamma(c-a-b)}{\Gamma(c-a)\Gamma(c-b)}
+C_1\left(\frac{l\sqrt{p}}{r}\right)^{d-1-2\beta}\frac{\Gamma(c)\Gamma(a+b-c)}{\Gamma(a)\Gamma(b)}~.
\end{equation}
On the other hand, when $r\rightarrow\infty$, Klein-Gordon equation approximates to
\begin{equation}\label{radiald}\
\partial_ {r}^2R(r)+\frac{d}{r}\partial_ {r}R(r)+\left(\frac{\omega^2l^2}{r^4}-\frac{Ql^2}{r^4}-\frac{m^2l^2}{r^2}\right)R(r)=0~.
\end{equation}
The solution of Eq.~(\ref{radiald}) is a linear combination of the Bessel function \cite{M. Abramowitz}\ given by
\begin{equation}
R(r)=\left(\frac{\sqrt{A}}{2r}\right)^{\frac{d-1}{2}}\left[D_1\Gamma{(1-C)}J_{-C}\left(\frac{\sqrt{A}}{r}\right)+D_2\Gamma{(1+C)}J_{C}\left(\frac{\sqrt{A}}{r}\right)\right],
\end{equation}
where
\begin{equation}
A=l^2(l^2\omega^2-Q)~,
\end{equation}
\begin{equation}
C=\frac{1}{2}\sqrt{(d-1)^2+4m^2l^2}~.
\end{equation}
Now, using the expansion of the Bessel function \cite{M.
Abramowitz}\ we find the asymptotic solution in the polynomial
form
\begin{equation}\label{Rasympd}\
R_{asymp.}(r)=D_1\left(\frac{\sqrt{A}}{2r}\right)^{\frac{d-1}{2}-C}+D_2\left(\frac{\sqrt{A}}{2r}\right)^{\frac{d-1}{2}+C},
\end{equation}
for $\frac{\sqrt{A}}{r}\ll1$. Introducing,
\begin{equation}
\widehat{D}_1\equiv
D_1\left(\frac{\sqrt{A}}{2}\right)^{\frac{d-1}{2}-C},\,\,\,
\widehat{D}_2\equiv
D_2\left(\frac{\sqrt{A}}{2}\right)^{\frac{d-1}{2}+C}~,
\end{equation}
we write Eq.~(\ref{Rasympd}) as
\begin{equation}\label{Rasympd1}\
R_{asymp.}(r)=\widehat{D}_1\left(\frac{1}{r}\right)^{\frac{d-1}{2}-C}+\widehat{D}_2\left(\frac{1}{r}\right)^{\frac{d-1}{2}+C}~.
\end{equation}

Comparison of Eqs.~(\ref{Rinfinity}) and (\ref{Rasympd1}),
regarding $\beta = \beta_-$, it  allows us to immediately  read off
the coefficients $\widehat{D}_1$ and $\widehat{D}_2$
 which can be decomposed in terms of the incoming and
the outgoing coefficients $D_{\mbox{\tiny in}}$ and
$D_{\mbox{\tiny out}}$, Refs. (\cite{Gonzalez:2010vv,
Gonzalez:2010ht, Birmingham:1997rj,Kim:1999un, Oh:2008tc,
Kao:2009fh}). Defining, $\widehat{D}_1=D_{\mbox{\tiny
in}}+D_{\mbox{\tiny out}}$ and $\widehat{D}_2=ih(D_{\mbox{\tiny
out}}-D_{\mbox{\tiny in}})$. In this way, the reflection and
transmission coefficients
 are given by
\begin{equation}
\Re=\frac{\left|D_{\mbox{\tiny
out}}\right|^2}{\left|D_{\mbox{\tiny in}}\right|^2}~,\label{coef1}
\end{equation}
\begin{equation}
\mathfrak{U}=\frac{\omega
l^{d-1}p^{\frac{d-4}{2}}\left|C_{1}\right|^2}{2\left|h\right|C\left|D_{\mbox{\tiny
in}}\right|^2}~,\label{coef2}
\end{equation}
and the absorption cross section, $\sigma_{abs}$, is given by
\begin{equation}\label{absorptioncrosssection}\
\sigma_{abs}=\frac{\mathfrak{U}}{\omega}=\frac{l^{d-1}p^{\frac{d-4}{2}}\left|C_{1}\right|^2}{2\left|h\right|C\left|D_{\mbox{\tiny
in}}\right|^2}~,
\end{equation}
where, the coefficients $D_{\mbox{\tiny in}}$ and $D_{\mbox{\tiny
out}}$ are given by
\begin{equation}\label{D11}\
 D_{\mbox{\tiny in}}= \frac{C_1}{2}\left[\left(l\sqrt{p}\right)^{2\beta_-}\frac{\Gamma(c)\Gamma(c-a-b)}{\Gamma(c-a)\Gamma(c-b)}+\frac{i}{h}\left(l\sqrt{p}\right)^{d-1-2\beta_-}\frac{\Gamma(c)\Gamma(a+b-c)}{\Gamma(a)\Gamma(b)}\right],
 \end{equation}
\begin{equation}\label{D21}\
 D_{\mbox{\tiny out}}=
 \frac{C_1}{2}\left[\left(l\sqrt{p}\right)^{2\beta_-}\frac{\Gamma(c)\Gamma(c-a-b)}{\Gamma(c-a)\Gamma(c-b)}-\frac{i}{h}\left(l\sqrt{p}\right)^{d-1-2\beta_-}\frac{\Gamma(c)\Gamma(a+b-c)}{\Gamma(a)\Gamma(b)}\right].
 \end{equation}

\begin{figure}[!h]
\begin{center}
\includegraphics[width=4.0in,angle=0,clip=true]{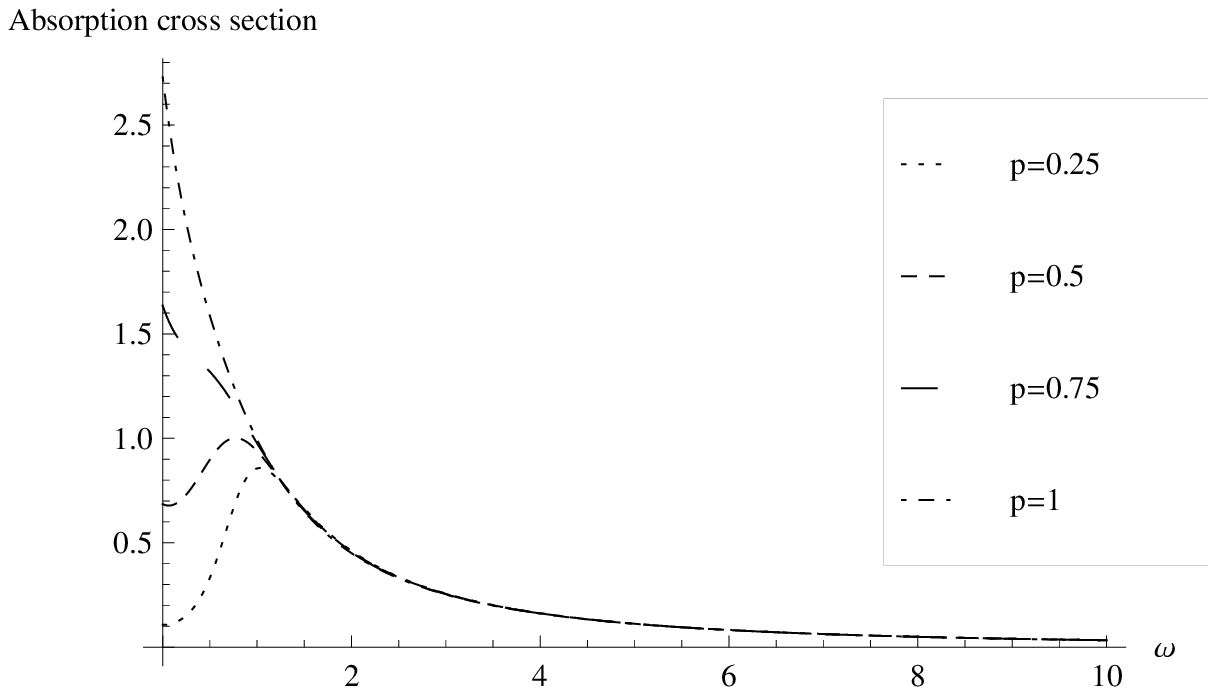}
\caption{Absorption cross section v/s $\omega$; $d=5$, $m^2l^2=-15/4$, $l=1$, $h=-1$ $\alpha=-1$ and $\xi=0$.}
\label{AbsorptionCrossSectionCSBH5dpxi0a}
\end{center}
\end{figure}

\begin{figure}[!h]
\begin{center}
\includegraphics[width=4.0in,angle=0,clip=true]{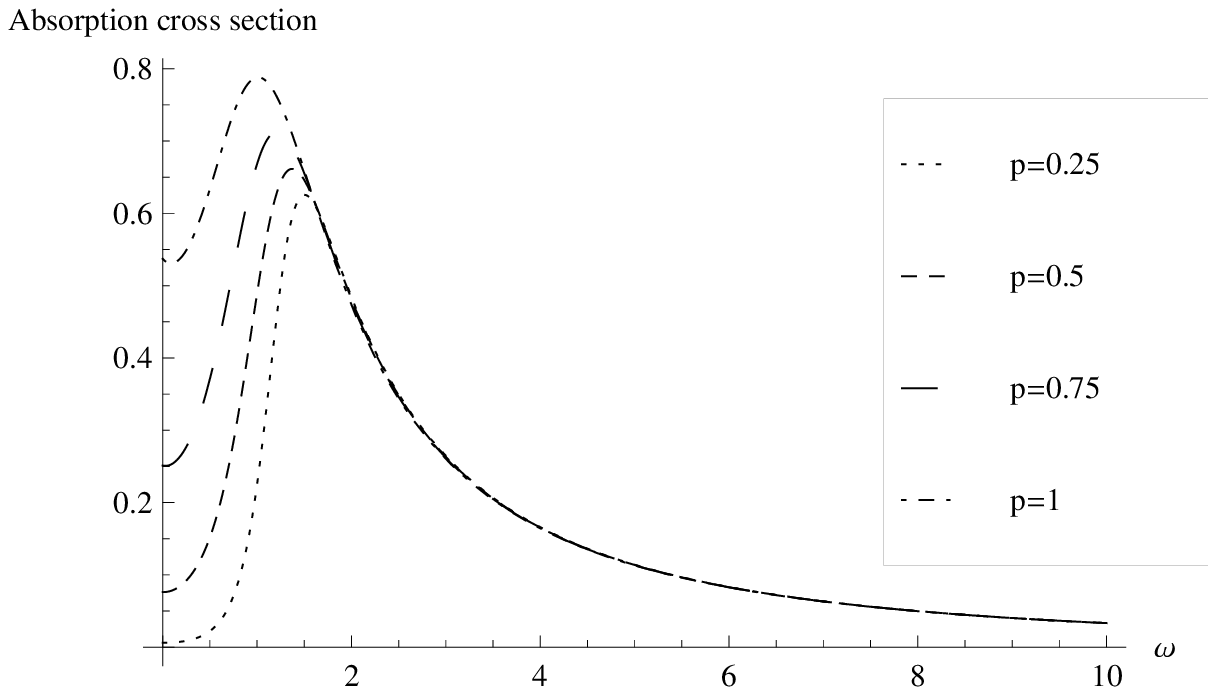}
\caption{Absorption cross section v/s $\omega$; $d=5$, $m^2l^2=-15/4$, $l=1$, $h=-1$, $\alpha=-1$ and $\xi=1$.}
\label{AbsorptionCrossSectionCSBH5dpxi1a}
\end{center}
\end{figure}

\begin{figure}[!h]
\begin{center}
\includegraphics[width=4.0in,angle=0,clip=true]{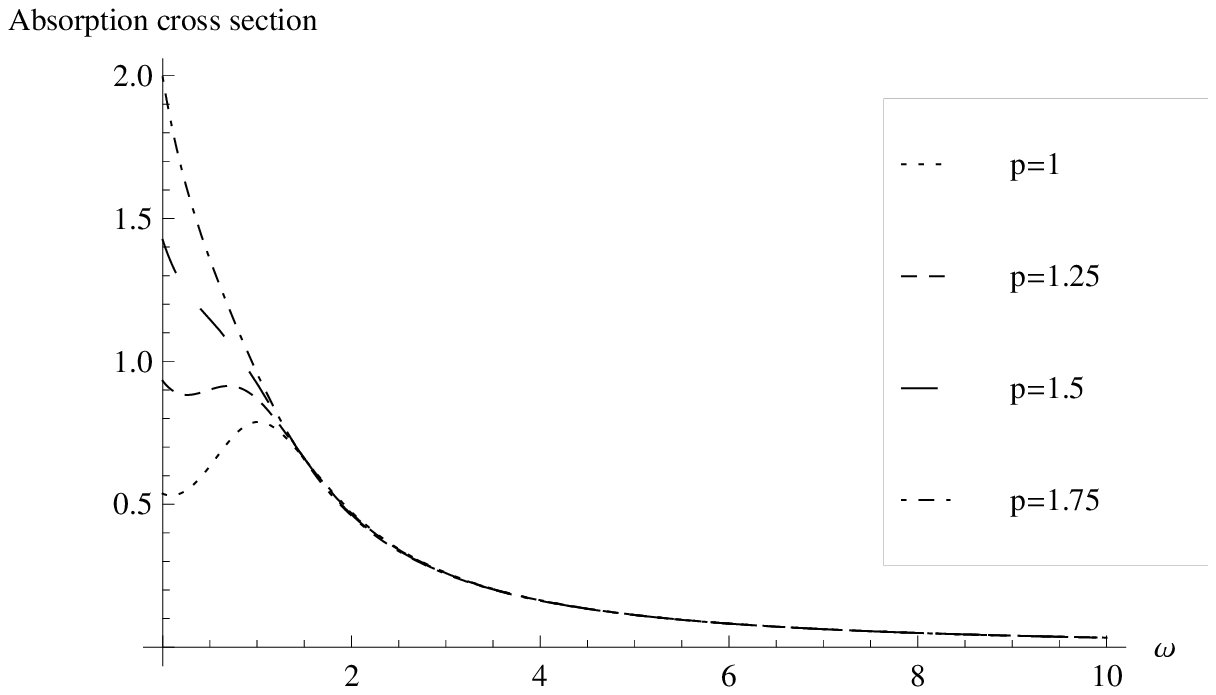}
\caption{Absorption cross section v/s $\omega$; $d=5$, $m^2l^2=-15/4$, $l=1$, $h=-1$, $\alpha=1$ and $\xi=1$.}
\label{AbsorptionCrossSectionCSBH5dpxi1b}
\end{center}
\end{figure}

\begin{figure}[!h]
\begin{center}
\includegraphics[width=4.0in,angle=0,clip=true]{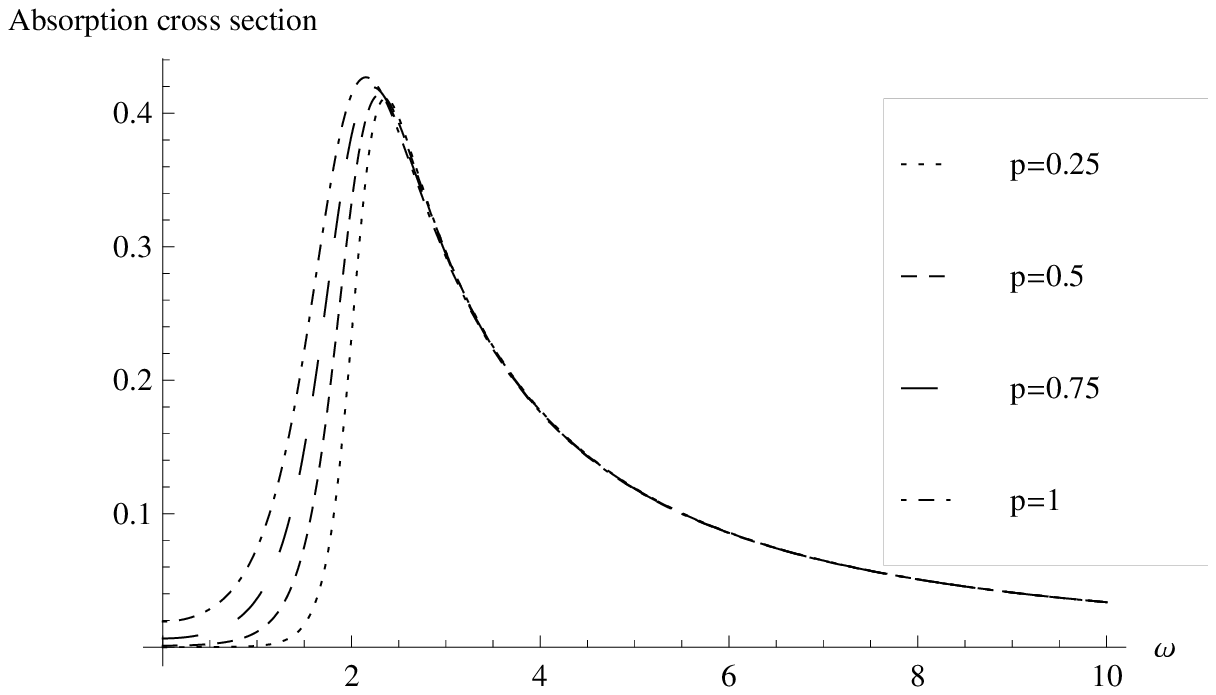}
\caption{Absorption cross section v/s $\omega$; $d=5$, $m^2l^2=-15/4$, $l=1$, $h=-1$, $\alpha=-1$ and $\xi=2$.}
\label{AbsorptionCrossSectionCSBH5dpxi2a}
\end{center}
\end{figure}

\begin{figure}[!h]
\begin{center}
\includegraphics[width=4.0in,angle=0,clip=true]{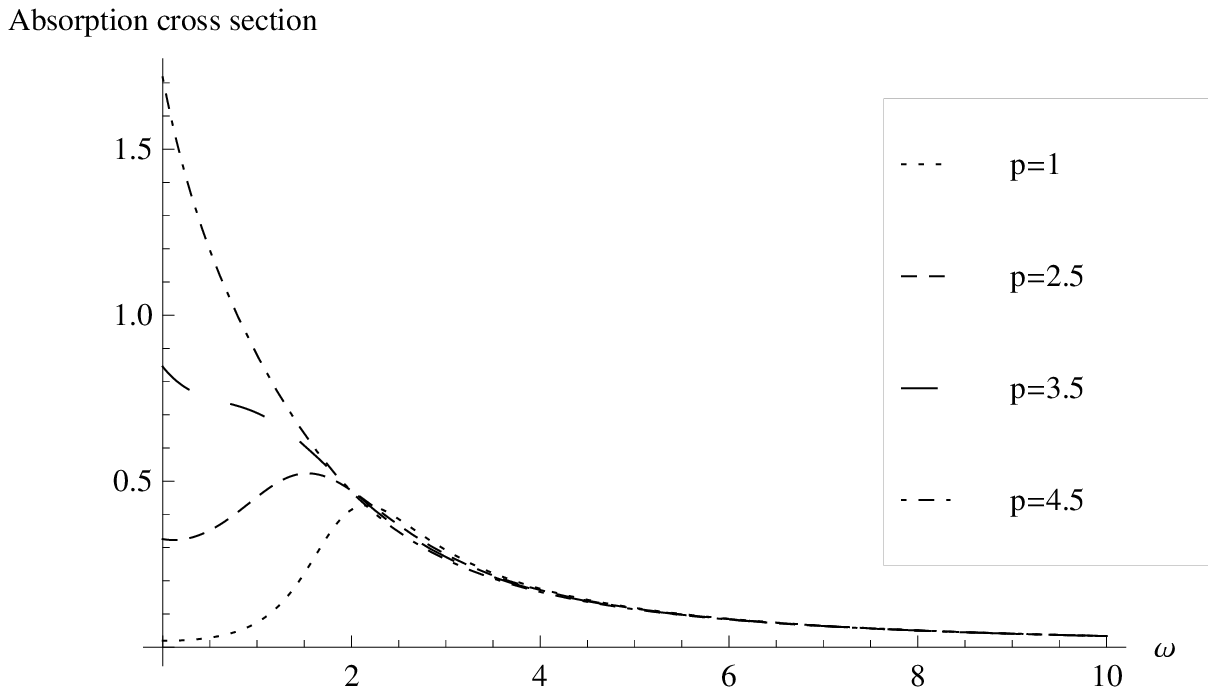}
\caption{Absorption cross section v/s $\omega$; $d=5$, $m^2l^2=-15/4$, $l=1$, $h=-1$, $\alpha=1$ and $\xi=2$.}
\label{AbsorptionCrossSectionCSBH5dpxi2b}
\end{center}
\end{figure}

Now, we will carry out a numerical analysis of the absorption
cross section given by the Eq.~(\ref{absorptioncrosssection}), for
a five-dimensional topological Chern-Simons black hole, $k=2$. We
consider without loss of generality, $m^2l^2=-15/4$, $l=1$ and we
fix $h=-1$. Then, we analyze the behavior of the absorption cross
section for various values of $p$, which it is given by
$p=1+\alpha(2\mu G_{k})^{\frac{1}{k}}$ with $\mu =\frac{\Omega
_{d-2}}{\Sigma _{d-2}}M$, for  a topological black holes. The two branches of
black hole solutions are given by $\alpha=\pm 1$. Therefore, for $k=2$ and
$\alpha=1$, then $\mu \geq0$ and $p\geq 1$ and for $k=2$ and
$\alpha=-1$, then $\frac{1}{2G_k}\geq\mu\geq0$ and $0\leq p\leq
1$. Massless black hole corresponds to $p=1$. Besides, we
restricted values of  $p$ and $\xi$, such as the quasinormal modes
exhibited real and imaginary part, i.e. $p<\xi^2+1$. Thus, we are
leaving out of our analysis the pure damping modes. Our results
for $\xi=0$ and $\alpha=-1$ are shown in
Fig.~(\ref{AbsorptionCrossSectionCSBH5dpxi0a}). Other branch
define by $\alpha=1$, it does not satisfy the restriction between
$\xi$ and $p$ described above. Then, for $\xi=1$ and $\alpha=\mp1$,
our results are shown in
Figs.~(\ref{AbsorptionCrossSectionCSBH5dpxi1a},
\ref{AbsorptionCrossSectionCSBH5dpxi1b}), respectively. Finally,
for $\xi=2$ and $\alpha=\mp1$  our results are shown in
Figs.~(\ref{AbsorptionCrossSectionCSBH5dpxi2a},
\ref{AbsorptionCrossSectionCSBH5dpxi2b}), respectively. In
general, for the branch $\alpha=-1$, the absorption cross section
decreases if the mass of black hole increase,
Figs.~(\ref{AbsorptionCrossSectionCSBH5dpxi0a},
\ref{AbsorptionCrossSectionCSBH5dpxi1a},
\ref{AbsorptionCrossSectionCSBH5dpxi2a}), at the zero-frequency
limit. For the other branch, $\alpha=1$, the behavior is opposite,
the absorption cross section increases if the black hole mass
increase, Figs.~(\ref{AbsorptionCrossSectionCSBH5dpxi1b},
\ref{AbsorptionCrossSectionCSBH5dpxi2b}). Also, we can see from the figures that the values of the coefficients converge at the values of coefficients for the topological massless black hole. Therefore, we can see that beyond a certain frequency the mass  black hole
does not affect the absorption cross section.

\section{Summary}
In this work we have studied the absorption cross section for a massive
scalar field propagating in a $5$-dimensional topological Chern-Simons black
hole. In this sense, at the low-frequency limit, we analyzed  the effect of the mass
black hole. We considered the
two branches of black hole solutions $\alpha=\pm 1$ and we showed
that, for the branch $\alpha=-1$, the absorption cross section
decreases if the mass of black hole increase, at the
zero-frequency limit. For the other branch, $\alpha=1$, the
behavior is opposite, the absorption cross section increases if
the black hole mass increase. We would like to note that the absorption cross section shows two characteristic
behaviors. The first one, it shows a maximum value for the
absorption cross section near to zero frequency and the
second  it shows a maximum for a finite non-vanishing value. Besides, beyond a certain frequency  the mass
black hole does not affect  the absorption cross section.




\section*{Acknowledgments}

This work was supported by COMISION NACIONAL DE CIENCIAS Y
TECNOLOGIA through FONDECYT Grant 1090613 (JS). This work was also
partially supported by PUCV DI-123.713/2011 (JS).


\appendix

\end{document}